\documentclass[letterpaper,prd,10pt,twocolumn,showpacs,showkeys,preprintnumbers,superscriptaddress,amsmath,amssymb,nofootinbib,notitlepage]{revtex4-1}
\usepackage{graphicx, epsfig, amssymb} %include figure files
\usepackage{amsmath, amsfonts}
\usepackage{bm} %include bold math: \bm{} creates bold letters in math mode
\usepackage{color}
\usepackage[usenames]{xcolor}
\usepackage{hyperref}
\usepackage{siunitx}
\hypersetup{linktocpage,colorlinks=true,urlcolor=blue,linkcolor=blue,citecolor=blue}  
\usepackage{verbatim}
\usepackage[utf8]{inputenc}
\usepackage{natbib}
\usepackage{bbm}
\usepackage{booktabs}

\def\rd{{\rm d}}
\def\rs{{\rm s}}

\begin{document}

\title{Gravitational Waves from Thurston Geometries}

\author{Daniel \surname{Flores-Alfonso}}
\email[]{danflores@unap.cl}
\affiliation{Instituto de Ciencias Exactas y Naturales, Universidad Arturo Prat, Avenida Playa Brava 3256, 1111346, Iquique, Chile}
\affiliation{Facultad de Ciencias, Universidad Arturo Prat, Avenida Arturo Prat Chac\'on 2120, 1110939, Iquique, Chile}

\author{Cesar S.~\surname{Lopez-Monsalvo}}
\email[]{cslm@azc.uam.mx}
\affiliation{Departamento de Ciencias B\'asicas, Universidad Aut\'onoma Metropolitana Azcapotzalco,  Avenida San Pablo No. 420 Colonia Nueva el Rosario C.P. 02128, Alcald\'ia Azcapotzalco, Ciudad de M\'exico, Mexico}

\author{Marco \surname{Maceda}}
\email[]{mmac@xanum.uam.mx}
\affiliation{Departamento de F\'isica,
Universidad Aut\'onoma Metropolitana - Iztapalapa,
Avenida Ferrocaril San Rafael Atlixco 186, Col. Leyes de Reforma 1ra Secci\'on, C.P. 09310, Ciudad de M\'exico, Mexico}


\begin{abstract}
It has been established that the famous three-dimensional Thurston geometries have four intrinsically Lorentzian analogs. We explore these spacetimes in three-dimensional general relativity nonminimally coupled to a scalar field together with electromagnetic matter. We find that three of these spacetimes support electromagnetic radiation, while, the other is partially sourced by a nonnull field and supports gravitational radiation. By addressing this problem we have also found a novel type of gravitational Cheshire effect. 
\end{abstract}

\maketitle

\section{Introduction}

The eight Thurston geometries are homogeneous spaces with positive-definite metrics that uniquely characterize every three-dimensional manifold. In cosmology, they have been used to build spatially homogeneous anisotropic models, see Refs.~\cite{Taub:1950ez,Kantowski:1966te,Fagundes:1985}. However, they also admit the interpretation of three-dimensional gravitational instantons. Along this line, in Ref.~\cite{Flores-Alfonso:2021opl} they were all shown to be vacuum solutions of three-dimensional massive gravity, see Ref.~\cite{Bergshoeff:2009hq}.

Three-dimensional Lorentzian spacetimes do not play the same fundamental role as the Thurston models\footnote{See, for instance, Theorem 5.8 of Ref.~\cite{Isham:1999}}. However, it is also true that some feats of Lorentzian geometries are impossible for metrics with Euclidean signature. Such a complementation is found in the work of Dumitrescu and Zeghib (DZ), Ref.~\cite{Dumitrescu:2010}, where they establish that there are only four spacetimes with uniquely Lorentzian properties that analog the Thurston geometries\footnote{Part of the proof in that work is corrected in Ref.~\cite{Allout:2022}.}. These three-dimensional geometries are given by the Minkowski spacetime, the Anti-de Sitter (AdS) space, a plane-wave spacetime and a homogeneous anisotropic universe; see Eq. \eqref{DZ} below. Incidentally, the maximally-symmetric spaces may be obtained by applying a Wick rotation to a Thurston geometry. However, that is certainly not the case for the other two geometries which are time-dependent Kundt spacetimes. 

Let us remark that unlike other homogeneous plane waves (see Ref.\cite{Torre:2012hw}) the DZ gravitational waves are not an electrovacuum spacetime. This is can be established with the methods of Ref.~\cite{Krongos:2016bqp}. Moreover, in order to prove this for all the DZ spacetimes we have extended that method in Appendix \ref{MCS}. Nonetheless, the DZ geometries have all been found to represent vacuum solutions of three-dimensional massive gravity, Ref.~\cite{Flores-Alfonso:2021opl}. That is, with the caveat that the plane waves solve the equations only in the massless limit. In such a case, the equations of motion are not an extension of Einstein's. Thus, this elusive gravitational wave has escaped full description within general relativity. We fill this gap by finding appropriate matter sources. 

The maximally-symmetric DZ metrics are, of course, of physical relevance and need no further discussion here. The other two are analogs of the Nil and Sol Thurston geometries; so-called because their Killing algebras are nilpotent and solvable, respectively. Those two Thurston spaces are actually three-dimensional Lie groups, the first of which is the Heisenberg group whose Lie algebra appears in the cannonical commutation relations of quantum mechanics. Both nilpotent and solvable isometries arise in Bianchi cosmologies of type $II$ and type $VI_0$, respectively. Moreover, in higher-dimensional settings, black holes have been constructed which also possess such isometries, see Refs.~\cite{Cadeau:2000tj,Hervik:2003vx,Hervik:2007zz,Hassaine:2015ifa,Bravo-Gaete:2017nkp}. The DZ spacetimes with Nil and Sol isometries and higher-dimensional versions of them have been studied in a number of different scenarios from string theory to thermodynamic fluctuation theory~\cite{Papadopoulos:2002bg,Blau:2002js,Blau:2004yi,Bravetti:2015,Lopez:2021,Flores-Alfonso:2020zif}.

In this work, we find that the DZ spacetimes are all solutions of three-dimensional general relativity nonminimally coupled to a scalar field and supplemented by electromagnetic matter. It is well-known that when one departs from minimal couplings then fields may no longer curve spacetime~\cite{Ayon-Beato:2005yoq}. This gravitational effect has also been reported for composite matter sources such as our own~\cite{Cardenas:2014kaa}. However, to the best of our knowledge, in this paper we present the first instance of this situation that incorporates null electromagnetic fields. In particular, this provides a new type of gravitational Cheshire effect.

We have organized our manuscript as follows:  In Sec. \ref{sec:II}, we write down the action principle we consider and its corresponding equations of motion. Subsequently, in Sec. \ref{sols} we present two sets of solutions: first considering matter fields which are invariant under the isometries of the spacetime metric and then abandoning this restriction. Thus finding solutions in three-dimensional gravity with nontrivial scalar and electromagnetic fields for each DZ spacetime. In the Appendix, we have included several supplementary material. Since the DZ geometries are left-invariant Lorentzian metrics on Lie groups we summarize their structure and relevant properties in Appendix \ref{groups}. In order to show the DZ spaces are, in fact, not three-dimensional electrovacuum spacetimes we provide a generalization of the three-dimensional Rainich conditions for the inclusion of a Chern-Simons term in Appendix \ref{MCS}. Lastly, we find it relevant to discuss the higher-dimensional versions of the DZ spacetimes as they have already been previously studied in the literature. Thus, in Appendix \ref{highD} we show how they solve the equations of motions and in which cases the theory must be modified in order to accommodate the higher dimension.

\section{Theory and Setup}
\label{sec:II}

Our problem is set up as follows: We have four specified spacetime geometries, namely
\begin{subequations} \label{DZ}
\begin{align}
\rd s_{\text{Mink}}^2 &= -2\rd u \rd v +\rd x^2,\\
\rd s^2_{\text{AdS}} &= \frac{1}{x^2}\left(-2\rd u \rd v +\rd x^2\right), \\
\rd s_{\text{Sol}}^2 &= \frac{2x^2}{u^2}\rd u^2-2\rd u \rd v +\rd x^2, \\
\rd s_{\text{Nil}}^2 &= -2\rd u \rd v +(\rd x+v\rd u)^2,
\end{align}    
\end{subequations}
and search for matter fields which source them in accordance with Einstein's equations. Almost a century ago, in Ref.~\cite{Rainich:1925}, Rainich showed what were the necessary and sufficient geometric conditions which a spacetime metric must satisfy in order for it to solve the four-dimensional Einstein-Maxwell equations together with some nonnull electromagnetic field. Much more recently, the three-dimensional Rainich conditions for null and nonnull fields were established in Ref.~\cite{Krongos:2016bqp}. These approaches are ideal for the problem at hand.

As it turns out, none of the DZ geometries satisfy these conditions completely. The subtleties present are as follows. Since the conditions separate according to whether the supported field is null or not, so do the DZ geometries. The Minkowski, AdS and Sol geometries could, in principle, support null fields, whereas, the Nil spacetime could potentially be sourced by a nonnull field.
Since the Minkowski and AdS spacetimes are Einstein manifolds, the electromagnetic field which is constructed from the conditions is vanishing. The Sol metric satisfies all but one of the conditions. As a result, the reconstructed field is imaginary. The Nil geometry does not satisfy the conditions presented in Ref.~\cite{Krongos:2016bqp}. However, when they are generalized to account for a Chern-Simons term then all but one of the conditions are satisfied. As above, the corresponding electromagnetic field fails to be real.

In three spacetime dimensions, electromagnetic fields are dual to scalar fields. This allowed for the (2+1)-dimensional Rainich conditions to be written down originally. However, the same fact discards the possibility of adding scalar fields to answer the problem we set out to solve. Nonetheless, this duality is obstructed when one considers nonminimal coupling. 

It has long been known that in flat spacetime conformally coupling a scalar field improves a theory's renormalizability properties, see Refs.~\cite{Callan:1970ze,Deser:1970hs}. Such a strategy also led to the first hairy black hole, Ref.~\cite{Bekenstein:1974sf}, one not necessarily determined in full by its mass, charge and angular momentum. Much more recently, in cosmology, inflation was famously shown to arise from nonminimally coupling a Higgs field to gravity, see Ref.~\cite{Bezrukov:2007ep}. Presently, nonminimally coupled scalar fields are also considered in quantum field theory on curved spacetimes to generalize early work on cosmological particle creation, see Refs.\cite{Barbado:2018qod,Barbado:2021wnn} and references therein. Moreover, they have been continuously used as alternatives or complements to higher-curvature theories, see  for example Refs.~\cite{Ayon-Beato:2010vyw,Ayon-Beato:2019kmz,Bravo-Gaete:2021kgt}. Hence, we find that opting for nonminimally coupled scalar fields is a physically reasonable approach.

Thus, we consider the following action functional
\begin{align} \label{action}
S[g,\Phi,A] = \frac{1}{2}&\int\rd^3x\sqrt{-g}(R-2\Lambda -\nabla\Phi^2\notag\\
&\qquad -\zeta R\Phi^2-m^2\Phi^2-F^2)
\notag\\
+\frac{\mu}{2}&\int A\wedge F,
\end{align}
where the field strength $F$ and the gauge potential $A$ are related through the definition $F=\rd A$. Here, $\Lambda$ is the cosmological constant and $\zeta$ is the coupling constant measuring nonminimal coupling of the scalar field to the background. The mass of the scalar and gauge fields are $m$ and $\mu$. We parameterize the scalar field mass as usual, using $m^2$. However, recall that this quantity need not be positive, see for instance Refs.~\cite{Breitenlohner:1982bm,Breitenlohner:1982jf}. Lastly, notice that for the gauge field, the mass term is topological.

\begin{subequations} \label{eom}
The equations of motion for the matter sector are
\begin{align}
\Box\Phi -\zeta R\Phi-m^2\Phi &= 0,
\label{eomPhi}\\
\rd \star F +\mu F&= 0,   
\end{align}
whereas the gravitational field complies with
\begin{align}
G_{\mu\nu}+\Lambda g_{\mu\nu} = 
&\nabla_{\mu}\Phi\nabla_{\nu}\Phi-\frac{1}{2}\nabla\Phi^2 g_{\mu\nu} -\frac{1}{2}m^2\Phi^2g_{\mu\nu}
\notag\\
&+\zeta\left( g_{\mu\nu}\Box-\nabla_{\mu}\nabla_{\nu}+G_{\mu\nu} \right)\Phi^2
\notag\\
&+ 2F_{\mu\alpha}F_{\nu}^{\phantom{\nu}\alpha}-\frac{1}{2}F^2g_{\mu\nu}.
\label{eomg}
\end{align}
\end{subequations}

\section{Solutions}
\label{sols}

In this section, we find matter fields which source each of the DZ spacetimes beginning with fields that respect the spacetime symmetries of the background supporting them. The Killing vectors of the maximally symmetric spaces are well known, while, for the other DZ geometries they are found in Appendix \ref{groups}.

In general, the simplest fields to consider are those which inherit the symmetries of the spacetime supporting them. Thus, we impose
\begin{equation} \label{inherit}
    {\cal L}_X \Phi =0, \qquad \text{and} \qquad {\cal L}_X F =0, 
\end{equation}
for any of the background's Killing vector fields $X$. As a consequence, the scalar field must be constant in all cases. 
Under such circumstances, the equations of motion are modified to
\begin{equation}
    m^2 = -\zeta R, \label{effR}
\end{equation}
and
\begin{equation}
    \left(1-\zeta\Phi_0^2\right)G_{\mu\nu}+\Lambda_{\text{eff}}g_{\mu\nu} = 2F_{\mu\alpha}F_{\nu}^{\phantom{\nu}\alpha}-\frac{1}{2}F^2g_{\mu\nu}, \label{effEM}
\end{equation}
for some constant scalar field $\Phi=\Phi_0$. In Eq. \eqref{effEM} we have defined $\Lambda_{\text{eff}}=\Lambda+\frac{1}{2}m^2\Phi_0^2$.

Firstly, let us point out that there is a singular point whenever, $1-\zeta\Phi_0^2=0$. Secondly, Eq. \eqref{effR} now represents a stringent geometrical constraint. For instance, cyclic-symmetric configurations, such as rotating black holes, satisfy the constraint only if they have neutral charge (cf Ref.~\cite{Martinez:1999qi}). In other words, nonvanishing electromagnetic fields are not always a possibility in such cases. Nonetheless, this does not represent a problem for the DZ geometries as they all have constant scalar invariants, see Ref.~\cite{Coley:2007ib}. Lastly, Eq. \eqref{effEM} shows under these conditions the theory effectively becomes that of Einstein-Maxwell. Thus, the standard Rainich conditions are altered in order to reconstruct the electromagnetic field. Ultimately, the addition of a constant scalar field now allows for real-valued electromagnetic fields on the Sol and Nil geometries. 

For the maximally symmetric DZ geometries, symmetry inheritance implies the electromagnetic field is vanishing. For the sake of completeness, we summarize how they solve Eqs. \eqref{eom}. The Minkowski solution simply requires $\Lambda=m^2=0$, while, for the AdS solution we must have $m^2=6\zeta$ and $\Lambda+2\zeta\Phi_0^2=-1$. 

The Sol solution is
\begin{subequations} \label{sol}
    \begin{align}
    \rd s ^2 &=\frac{2x^2}{u^2}\rd u^2-2\rd u \rd v +\rd x^2,\\
    \Phi &= \Phi_0,\\
    F &= \frac{\sqrt{\zeta\Phi_0^2-1}}{u}\rd u \wedge \rd x,
    \end{align}
\end{subequations}
for which we require $\Lambda=m^2=0$, the Minkowski conditions, together with $\mu=0$ and $\zeta\Phi_0^2-1>0$. In this system, electromagnetic waves travel aligned with the gravitational waves. Since the background is a conformally flat plane-wave spacetime then all of its curvature is given by the Ricci tensor. Moreover, since the wave profile in the metric is positive this causes the Ricci curvature to have the form of a pure-radiation tensor with opposite sign. In particular, this shows that the waves cannot be sourced purely by electromagnetic radiation. Furthermore, since the Ricci tensor couples to the square of the constant scalar field this causes the contribution of the scalar field to also have pure-radiation form with opposite sign.

The Nil solution is
\begin{subequations} \label{nil}
    \begin{align}
    \rd s ^2 &=-2\rd u \rd v +(\rd x+v\rd u)^2,\\
    \Phi &= \Phi_0,\\
    F &= \sqrt{\frac{\zeta\Phi_0^2-1}{2}} \rd u \wedge \rd v,
    \end{align}
\end{subequations}
and it requires $\Lambda=\frac{1}{4}$, $m^2=-\frac{1}{2}\zeta$ and $\mu=1$, as well as, $\zeta\Phi_0^2-1>0$. This configuration, has a homogeneous electromagnetic field making it quite similar to the Bertotti-Robinson and Pleba\'nski-Hacyan electrovacuum universes, see Refs.~\cite{Bertotti:1959pf,Robinson:1959ev,Plebanski:1979}. The background belongs to the Kundt family of geometries, a large class which generalizes the pp-wave spacetimes. It admits two independent null geodesic non-expanding congruences which define a cannonical Newman-Penrose triad, see Ref.~\cite{Flores-Alfonso:2020zif}. The comoving observer defined by the triad experiences a constant nonvanishing shear yet no expansion or rotation. The electromagnetic field is aligned with the triad, however, the field itself is not null. As with the plane waves above, the total contribution of the matter sector behaves effectively as an electromagnetic field with the opposite sign in the energy-momentum tensor. Once more, this is due to the nonminimal coupling of the scalar field.

In both cases, the electromagnetic field is dual to the central element of their Killing Heisenberg subalgebra $\xi$; see Eqs. \eqref{xisol} and \eqref{xinil}. Recall that this vector field is null for the Sol metric as it is tangent to its null geodesic congruence. For the Nil spacetime it is decisively spacelilke which in this context represents a nonnull electromagnetic field. This duality implies the electromagnetic fields are covariantly constant along arbitrary vector fields parallel to $\xi$. For the Sol solution the field is also covariantly preserved along arbitrary vector fields parallel to $\frac{\partial}{\partial x}$, as it would be for any other \emph{pp} wave. Lastly, let us note that the Sol and Nil solutions require both a scalar field and a gauge field, for example, in the limit where the gauge field vanishes Eq. \eqref{effEM} becomes singular.

To close this section, we investigate travelling waves on the Nil metric. By which is meant that we construct a Kerr-Schild spacetime with geometry described by
\begin{subequations} \label{nilwaves}
\begin{equation}
    \rd s ^2 =2H(u,x)\rd u^2-2\rd u \rd v +(\rd x+v\rd u)^2,
\end{equation}
together with the matter content in Eq. \eqref{nil}, for further reading on such metrics consult Refs.~\cite{Kerr:1965,Taub:1981evj}. This configuration solves Eqs. \eqref{eom} whenever 
\begin{equation}
H(u,x) = F_1(u)+F_2(u)e^{-x}.    
\end{equation}
\end{subequations}
Of course, the Nil conditions must be satisfied as well, i.e., $\Lambda=\frac{1}{4}$, $m^2=-\frac{1}{2}\zeta$, $\mu=1$ and $\zeta\Phi_0^2-1>0$. These gravitational waves are similar to the electrovacuum waves studied in Refs.~\cite{Plebanski:1979,Garcia:1984,Khlebnikov:1986sw,Garfinkle:1992dz,Ortaggio:2002bp,Podolsky:2002sy}.

\subsection{Less restrictive solutions}

In the previous section, we discussed one way in which the Minkowski and AdS spacetimes are solutions of Eq. \eqref{eom}. The matter content in those solutions inherit the symmetries of the background supporting them; they satisfy Eqs. \eqref{inherit}. As a consequence, the gauge field vanishes. We now dispose of such a restriction in order to find solutions with a nonvanishing electromagnetic field. Stealth, as defined in Ref.~\cite{Ayon-Beato:2004nzi}, is a noteworthy manner in which we rely on to construct such fields. A supreme example of stealth is provided by the gravitational Cheshire effect, in which nontrivial fields go undetected by flat spacetime (see Ref.~\cite{Ayon-Beato:2005yoq}).

By their very nature, stealth fields may forgo any sort of spacetime symmetry inheritance. On the other hand, fields representing pure radiation very closely resemble stealth fields in that they solve very nearly the same set of equations of motion. Such similarity was kept in mind in Ref.~\cite{Ayon-Beato:2005gdo} where \emph{pp}-wave spacetimes sourced by nonminimally coupled matter were analyzed. In this work, we follow suit as the Rainich conditions show that the Minkowski, AdS and Sol spacetimes have adequate geometries for supporting electromagnetic radiation.

Let us start by writing down our Ansatz as
\begin{subequations} \label{AnsatzM}
    \begin{align}
    \rd s ^2 &=-2\rd u \rd v +\rd x^2,\\
    \Phi &= \Phi(u),\\
    F &= f(u) \rd u \wedge \rd x,
    \end{align}
\end{subequations}
for which Eqs. \eqref{eom} are satisfied when $\Lambda=m^2=\mu=0$ and
\begin{equation} \label{PhiEq}
    2\zeta\Phi\Phi''-(1-2\zeta)(\Phi')^2=2f(u)^2. 
\end{equation}
Once the profile of the radiation field is specified this equation may be solved for the scalar field. Indeed, by making the redefinition $f(u)=h(u)\Phi'$ then Eq. \eqref{PhiEq} becomes
\begin{equation}
    2\zeta\frac{(\Phi\Phi')'}{(\Phi')^2}-1=2h(u)^2,
\end{equation}
whose solution is
\begin{equation} \label{PhiM}
    \Phi=c_2\exp{\left[-2\xi\int \frac{\rd u}{c_1+(1-4\xi)u+2\int h(u)^2\rd u  }\right]}.
\end{equation}

These solutions constitute a type of gravitational Cheshire effect. In this version, radiation propagates through Minkowski spacetime without curving it due to the presence of a scalar field that couples to the background in such a way that it exactly cancels the electromagnetic energy-momentum tensor. Let us recall that the standard Cheshire effect allows for a self-interaction potential. In the present case, the Maxwell field is null and does not play an effective substitution of the self-interacting potential. It is genuinely a novel kind of gravitational effect.

Notice that when one considers flat spacetime with trivial fields the value of the on-shell action vanishes, cf. Eq. \eqref{action}. This is unsurprising as this is a very basic configuration of the theory. However, for fields given by Eq. \eqref{AnsatzM} the action also vanishes
\begin{equation}
    S_{\text{on-shell}}=0.
\end{equation}
This happens because the Maxwell field is null and the scalar field is perceived by the action as pure radiation. Thus, the solutions we have found which constitute a new type of Cheshire effect are energetically on par with the most basic flat space solution.

We also remark on the fact that half of the spacetime symmetries are not themselves symmetries of the fields. Curiously enough, the fields do inherit the symmetries of the Killing Heisenberg subalgebra conformed by
\begin{subequations} \label{heisM}
\begin{align}
    P&=\frac{\partial }{\partial x},\\
    Q&=x\frac{\partial }{\partial v}+u\frac{\partial }{\partial x},\\
    \xi&=\frac{\partial }{\partial v}.
\end{align}
\end{subequations}

As it turns out, given a scalar field exhibiting stealth on some spacetime then it is possible to use it as a seed solution to generate new configurations via conformal transformations, see Refs.~\cite{Ayon-Beato:2013bsa,Ayon-Beato:2015mxf} and references therein. This is relevant to us since the Minkowski, AdS and Sol geometries are in the same conformal class.

Our next step is to establish AdS spacetime as a solution of Eqs. \eqref{eom} with a nonvanishing gauge field. For this we use the conformal factor relating AdS spacetime to flat space, i.e., $\Omega=\frac{1}{x}$. Indeed, by using Eq. \eqref{AnsatzM} as a seed solution we find that the configuration
\begin{subequations}
    \begin{align}
    \rd s ^2 &=\Omega^2\left(-2\rd u \rd v +\rd x^2\right),\\
    \Phi &= \Omega^{-\rs}\Phi_{\text{Mink}},\\
    F &= \Omega^{1-\rs}F_{\text{Mink}},
    \end{align}
also solves Eqs. \eqref{eom} and reduces to Eq. \eqref{PhiEq} where $\rs=\frac{2\zeta}{1-4\zeta}$. The configuration requires $\Lambda=-1$ and
    \begin{align}
        m^2 &= \frac{6\zeta(\zeta-1/8)(\zeta-1/6)}{(\zeta-1/4)^2}, \\
        \mu &= \frac{\zeta/2}{\zeta-1/4},
    \end{align}
\end{subequations}
meaning both the scalar and gauge field are massive. This is the same mass that was found for scalar fields in Ref.~\cite{Ayon-Beato:2005gdo}. Whereas, the gauge field has a mass simply given by the opposite of the parameter $\rs$. The scalar field may be massless in some cases, e.g., when it couples conformally to the background and so $\zeta=\frac{1}{8}$. However, the gauge field is always massive signaling that it is crucially non-Maxwellian. 

This solution represents electromagnetic waves travelling undetected over AdS spacetime. Nonminimal coupling of the free scalar field counterbalances the energy-momentum tensor carried by the waves. Regarding symmetry inheritance, the configuration is quite remarkable. Since the background is maximally symmetric its isometry group is six-dimensional. Yet the only symmetry inherited by the fields is translation along the null geodesic congruence.

To close this section, we explore non-inheriting fields not exhibiting stealth focusing on the special value $\zeta=\frac{1}{4}$. To begin, let us take a second look at the Killing subalgebra in Eq. \eqref{heisM}. If one thinks of the Minkowski spacetime as a flat plane wave then the subalgebra under consideration represents the Heisenberg symmetry characteristic of all plane-wave spacetimes, see Refs.~\cite{Allout:2022,Blau:2002js}. Since, the Sol geometry is also such a gravitational wave we consider the following Ansatz
\begin{subequations}
    \begin{align}
    \rd s ^2 &=\frac{2x^2}{u^2}\rd u^2-2\rd u \rd v +\rd x^2,\\
    \Phi &= \Phi(u),\\
    F &= f(u) \rd u \wedge \rd x,
    \end{align}
\end{subequations}
whose fields do not inherit the scaling symmetry associated with Eq. \eqref{T}.

Inserting our chosen Ansatz into Eq. \eqref{eom} yields
\begin{equation}
     \frac{\Phi^2}{4}\left(\log{\frac{\Phi}{u}}\right)''-\frac{1}{u^2}=f(u)^2,
\end{equation}
together with $\Lambda=m^2=\mu=0$. A particular solution is given by
\begin{equation}
    \Phi = u\exp{(ku+\sigma_0)}+\Phi_0, \label{nisol}
\end{equation}
where $k$, $\sigma_0$ and $\Phi_0$ are constants satisfying $k\geq0$ and $\Phi_0\geq2$, of which the latter is equivalent to $1-\zeta\Phi_0^2\geq0$. The limit $\sigma_0\to-\infty$ recovers the solution of the previous section evaluated at $\zeta=\frac{1}{4}$. A stealthy field is found outside of the allowed range, as it should, where $\Phi_0=0$.

\section{Closing Remarks}

In this work, we have have considered three-dimensional general relativity with an electromagnetic field together with a scalar field nonminimally coupled to gravity. Moreover, each of the fields is allowed to be massive. Our objective has been to find a matter model that is able to source each of the four spacetime geometries found by Dumitrescu and Zeghib in Ref.~\cite{Dumitrescu:2010}. Due to their properties these spacetimes are of mathematical interest, however, they have also been recently shown to be vacuum solutions of three-dimensional massive gravity, see Ref.~\cite{Flores-Alfonso:2021opl}. The geometries are all of Kundt type and include maximally symmetric spaces, as well as, a gravitational-wave spacetime. Moreover, metrics quite similar to them have previously arisen in the physics literature. Thus, we consider them to be of physical relevance. 

In summary, we have found self-gravitating systems whose background metric is given by one of the following: i) Minkowski spacetime ii) Anti-de Sitter space iii) time-dependent homogeneous plane waves or iv) a homogeneous universe with Bianchi II anisotropy. Searching for fields that themselves respect the spacetime symmetries of the background led us to the unique solutions within the theory, since the backgrounds are highly symmetric. For the maximally symmetric spacetimes this implies that the fields are trivial. Which is also expected given that they are Einstein spaces.

In order to fully probe the theory, we have also considered less constrained systems finding nontrivial fields over the maximally symmetric spacetimes. These fields are able to solve the equations of motion since their total energy-momentum tensor cancels itself out. Stealthy behavior of matter such as this is possible when scalar fields are nonminimally coupled to gravity, see Ref.\cite{Ayon-Beato:2004nzi}. In particular, our results provide a novel kind of gravitational Cheshire effect (Ref.\cite{Ayon-Beato:2005yoq}) in which electromagnetic radiation overflies flat space undetected due to a counteracting scalar field; an effect present in all spacetime dimensions. Although we focus primarily on the three-dimensional case we have also outlined how our results carry over to higher dimensions. Thus, although one can raise an argument on the physics of stealth fields, their mathematical properties provides us with a tool for generating new solutions via conformal transformations.

To close, let us mention that none of our targeted spacetimes are electrovacuum solutions. This is due to the nature of their curvature. This is to say, when one employs rigorous methods in order to reconstruct an electromagnetic field source from the geometry the result is either that the field vanishes or that it is imaginary. In our solutions, electromagnetic fields cannot act alone for this very reason. Moreover, in three-dimensions, standard scalar fields are dual to electromagnetic fields and so cannot resolve this issue. However, by nonminimally coupling scalar fields to gravity we have found a resolution. 

In most of our solutions, the gauge field is null thus representing electromagnetic waves. Hence, in order for the equations to be solved the scalar sector must cancel out this contribution and account for the spacetime curvature at the same time. However, if the electromagnetic field is turned off the scalar sector becomes singular. In other words, the scalar field cannot act alone either.

For the Einstein spaces, this cancelation occurs completely within the matter sector. This is to say, the standard contributions to the energy-momentum tensor are exactly canceled by the effects of nonminimal coupling to gravity. Consequently, the fields conform a composite stealth system. One in which neither of the fields exhibit stealth by themselves. Three-dimensional black holes with this behaviour were found in Ref.~\cite{Cardenas:2014kaa}. However, our solutions are the first instance of this occurring with electromagnetic radiation.

\section*{Acknowledgements}

We are grateful for helpful discussions with Eloy Ay\'{o}n-Beato, Crist\'{o}bal Corral, Mokhtar Hassa\"{i}ne, Mar\'{i}a Montserrat Ju\'{a}rez-Aubry and Pedro A. S\'{a}nchez. D.F.-A. thanks the members and staff of the Departamento de F\'{i}sica del Centro de Investigaci\'{o}n y de Estudios Avanzados for their hospitality while part of this manuscript was being written. His research is supported by Agencia Nacional de Investigaci\'{o}n y Desarrollo under FONDECYT grant No. 3220083.

\appendix

\section{Group and Geometric Structure}
\label{groups}

The DZ homogeneous spacetimes are the unique intrinsically Lorentzian maximal geometries which can be modeled on compact three-manifolds. The four of them are Lie groups with left-invariant metrics. Geometrically, they are all of Kundt type, meaning they possess a geodesic null congruence whose orthogonal distribution is integrable and represents a totally geodesic foliation of spacetime, see Refs.~\cite{Coley:2009ut,Boucetta:2022vny}. This structure is parameterized by a vector field $V$ that is tangent to the congruence and a one-form $\alpha$ such that
\begin{equation}
    \nabla_XV=\alpha(X)V,
\end{equation}
for all vector fields $X$ orthogonal to $V$. What is more, for three-dimensional Lie groups these structures are necessarily generated by left-invariant vector fields. Throughout, we use coordinates $\{u,v,x\}$ adapted to the geodesic null congruence so that it corresponds to translations along $v$. For all but one of the DZ geometries, these translations conform a Killing congruence as well, see the table just below.

\begin{center}
\vbox{\begin{equation*}\label{table1}
\renewcommand{\arraystretch}{1.5}
\begin{tabular}{cccc}
\toprule[1pt]\toprule[1pt]
 Spacetime & Group &  $\alpha$ &  Killing Congruence\\ 
 \midrule[0.5pt]
 Mink & E(3) & $\rd u$ & Yes \\ 
 \midrule[0.5pt]
 AdS & $\widetilde{\text{SL}}(2,\,\mathbb{R})$ & $-1/x\rd x$ & Yes \\
 \midrule[0.5pt]
 Sol & Sol & $\rd u$ & Yes \\
 \midrule[0.5pt]
 Nil & H(3) & $1/2\rd x$ & No \\
 \bottomrule[1pt]
\end{tabular}
\renewcommand{\arraystretch}{1}
\end{equation*}
}
\end{center}

Two of DZ geometries are irreducible symmetric spaces: the Minkowski and AdS spacetimes. The other two are left-invariant metrics on solvable Lie groups. On the Sol group the metric happens to be conformally flat. Lastly, on the Heisenberg group the metric is such that the center of its Killing algebra is a spacelike vector field.

The Minkowski metric
\begin{equation}
    \rd s^2 = -2\rd u \rd v +\rd x^2, 
\end{equation}
describes the unique intrinsically Lorentzian geometry which can placed on the Euclidean group. It is also supported by the Sol and Heisenberg groups, however, in the Sol case the geometry is not maximal; the details are described in Ref.~\cite{Dumitrescu:2010}. In the Heisenberg case, the center of its nilpotent algebra is a null vector field. However, the fact that such a metric is flat is a dimensional accident, as explained in Ref.~\cite{Vukmirovic:2015}.

In contrast, the universal cover of SL(2,\,$\mathbb{R}$) admits three distinct intrinsically Lorentzian geometries. However, among them only the one with constant sectional curvature (CSC) is maximal, i.e.,
\begin{equation}
    \rd s^2 = \frac{1}{x^2}\left(-2\rd u \rd v +\rd x^2\right). 
\end{equation}
Notice that when it comes to curvature, the AdS spacetime is a much more closely related to the hyperbolic Thurston geometry than to $\widetilde{\text{SL}}(2,\,\mathbb{R})$.  

The Sol geometry, described by
\begin{equation}
    \rd s^2 = \frac{2x^2}{u^2}\rd u^2-2\rd u \rd v +\rd x^2, 
\end{equation}
has one of seven left-invariant Lorentzian metrics on the Sol group, two of which are \emph{pp} waves. This list of metrics was reported in Ref.~\cite{Boucetta:2022}. The Sol metric describes plane waves travelling over a Minkowskian background and is the only such (non-flat) spacetime to admit compact models. Among the DZ geometries, only the Sol geometry is geodesically incomplete; a property that it shares with all non-unimodular homogeneous plane waves (cf. Ref.~\cite{Allout:2022}).

The Killing algebra of the Sol spacetime is generated by
\begin{subequations}
\begin{align}
    P&=\frac{2}{3}ux\frac{\partial }{\partial v}+\frac{1}{3}u^2\frac{\partial }{\partial x},\\
    Q&=\frac{x}{u^2}\frac{\partial }{\partial v}-\frac{1}{u}\frac{\partial }{\partial x},\\
    \xi&=\frac{\partial }{\partial v},\label{xisol}\\
    T&=u\frac{\partial }{\partial u}-v\frac{\partial }{\partial v}, \label{T}
\end{align}
\end{subequations}
of which $T$, $Q$ and $\xi$ are left-invariant fields satisfying the Sol algebra: $[\xi,Q]=0$, $[T,\xi]=\xi$ and $[T,Q]=-Q$. Its derived algebra locally represents the action of the Heisenberg group on all plane-wave spacetimes, more in depth discussions are found in Refs.~\cite{Allout:2022,Blau:2002js}. For the Sol plane wave
the Heisenberg algebra is extended by a scaling symmetry, cf. Eq. \eqref{T}. Gravitational waves with exactly these symmetries have been found to be exactly solvable string models (see Ref.~\cite{Papadopoulos:2002bg}) and to arise as the Penrose limit of a large class of spacetimes with singularities, for further details consult Refs.~\cite{Blau:2001ne,Blau:2002js,Blau:2003dz,Blau:2004yi}.

Along the same lines as above, the Heisenberg group admits a metric whose entire list of curvature scalars are constant. This is not obvious, as in the cases above, yet all constant scalar invariant (CSI) spacetimes are known in three dimensions from Ref.~\cite{Coley:2007ib}. Moreover, of the DZ spaces, only the Nil geometry
\begin{equation}
    \rd s^2 = -2\rd u \rd v +(\rd x+v\rd u)^2, 
\end{equation}
is not conformally flat. It is also the only instance in which the left-invariant vector field $V$ does not define a Killing congruence. Indeed, the Killing vector fields of the Nil spacetime are
\begin{subequations}
\begin{align}
    P&=\frac{\partial }{\partial u},\\
    Q&=\frac{\partial }{\partial v}+u\frac{\partial }{\partial x},\\
    \xi&=\frac{\partial }{\partial x},\label{xinil}\\
    T&=u\frac{\partial }{\partial u}-v\frac{\partial }{\partial v}.
\end{align}
\end{subequations}
In this case, $T$ is not a left-invariant vector field and the underlying Heisenberg algebra is $[P,Q]=\xi$, $[P,\xi]=0$ and $[Q,\xi]=0$. Notice that the central element $\xi$ is spacelike, as mentioned above. This is relevant as it was shown in Ref.~\cite{Rahmani:1992} that there are only three left-invariant metrics on the Heisenberg group and that distinguished by their central element, depending on whether that vector field is spacelike, timelike or null. %Pseudo-Riemannian manifolds such as the Nil spacetime surface in thermodynamics as phase spaces, as explained in Refs.\cite{Bravetti:2015,Lopez:2021}. In three dimensions, the phase space is Lorentzian and the canonical coordinates coincide with adapted Kundt coordinates. Although the phase spaces are not Lorentzian, they have been shown to solve the Einstein-Gauss-Bonnet equations and some spacetimes do emerge as equilibrium state geometries within them~\cite{Bravetti:2015ija}.

In three-dimensions, the Ricci and Cotton tensors together play a role similar to the Ricci and Weyl tensors in higher dimensions. In this case, the Riemann tensor is completely determined by the Ricci curvature and the Cotton tensor is traceless and conformally-invariant. Thus, topologically massive gravity (TMG) is a remarkable theory of gravity as its equations of motion relate these two tensors; details are described in Refs.~\cite{Deser:1981wh,Deser:1982vy}. For TMG vacua a non-ambiguous algebraic type is determined and a Goldberg-Sachs theorem applies~\cite{Nurowski:2015hwa}. Except for the Sol geometry, the DZ spacetimes are TMG vacua and so the theorem implies they are algebraically special. What is more, their algebraic type is $D$ or more special and so they possess two distinct divergence-free geodesic null congruences.

None of the DZ geometries represents an eletrovacuum spacetime of general relativity. However, the degree to which they fail to do so is a geometric measure, as determined by the Rainich conditions of Ref.~\cite{Krongos:2016bqp}. A combination of curvature scalars separates spaces into those which could support null or nonnull electromagnetic fields. The Nil metric satisfies the nonnull condition and is, in fact, an electrovacuum of TMG. As exemplified in Ref.~\cite{Flores-Alfonso:2020zif}, the Nil metric supports a nonnull Maxwell-Chern-Simons field. In order to fully exploit the three-dimensional version of the Rainich conditions we slightly generalize the results of Ref.~\cite{Krongos:2016bqp} in Appendix \ref{MCS} by considering the possibility of a Chern-Simons term. The remaining DZ spacetimes satisfy the null field condition, however, since they are conformally flat there is no hope of them describing TMG-electrovacua. In section \ref{sec:II}, we show that they all satisfy the Einstein equations when the background is nonminimally coupled to a scalar field and supports an electromagnetic field as well. A relevant property of these solutions is that the scalar curvature is constant. In the table below we display the Ricci scalar $R$ of each DZ metric future reference.

\vbox{
\begin{equation*}\label{table2}
\renewcommand{\arraystretch}{1.5}
\begin{tabular}{cccc}
\toprule[1pt]\toprule[1pt]
 Spacetime & Curvature & $R$ &  Conformally Flat\\ 
 \midrule[0.5pt]
 Mink & Flat & $0$ & Yes \\ 
 \midrule[0.5pt]
 AdS & CSC$^{a}$ & $-6$ & Yes \\
 \midrule[0.5pt]
 Sol & VSI$^{b}$ & $0$ & Yes \\
 \midrule[0.5pt]
 Nil & CSI$^{c}$ & $1/2$ & No \\
 \bottomrule[1pt]
\end{tabular}
\renewcommand{\arraystretch}{1}
\end{equation*}
}
\begin{flushleft}
\quad\qquad $^{a}$ constant sectional curvature\\
\quad\qquad $^{b}$ vanishing scalar invariants\\
\quad\qquad $^{c}$ constant scalar invariants
\end{flushleft}

\section{Rainich Conditions for Einstein-Maxwell-Chern-Simons theory}
\label{MCS}

The necessary and sufficient conditions for a (2+1)-dimensional spacetime to satisfy the Einstein-Maxwell equations,
\begin{equation}
    G_{\mu\nu} + \Lambda g_{\mu\nu} = 2F_{\mu\alpha}F_{\nu}^{\phantom{\nu}\alpha}-\frac{1}{2}F_{\alpha\beta}F^{\alpha\beta}g_{\mu\nu},
\end{equation}
for some field $F$ that also complies with the Maxwell-Chern-Simons equations
\begin{equation}
    \rd F=0, \qquad \text{and} \qquad \rd \star F+\mu F=0,
\end{equation}
are called the Rainich conditions. The conditions separate according to whether the field is null or not, i.e., $F_{\alpha\beta}F^{\alpha\beta} = 0$.

For a nonnull field it is instructive to define the following Einstein scalars
\begin{equation}
    G = G_{\alpha}^{\alpha}, \quad {}_2G = G_{\alpha}^{\beta}G_{\beta}^{\alpha}, \quad {}_3G = G_{\alpha}^{\beta}G_{\beta}^{\gamma}G_{\gamma}^{\alpha},
\end{equation}
and the tensor whose components are
\begin{equation}
    H_{\mu\nu} = G_{\mu\nu}-(G+2B)g_{\mu\nu},
\end{equation}
where
\begin{equation}
    B = \frac{1}{2}\frac{\frac{1}{3}G{}_2G-{}_3G}{{}_2G-\frac{1}{3}G^2}.
\end{equation}
The Rainich conditions are then
\begin{align}
{}_2G-\frac{1}{3}G^2 \neq 0,\\
H_{\alpha\beta}X^{\alpha}X^{\beta} >0, \\
B=\Lambda,\\
H_{\mu[\nu}H_{\rho]\sigma}=0,
\end{align}
for some vector field $X$ and
\begin{align}
     H_{\mu\nu}H_{\rho[\sigma;\tau]}+H_{\mu\rho}H_{\nu[\sigma;\tau]}+H_{\nu\rho[;\sigma}H_{\tau]\mu}\notag\\
     =-\mu \varepsilon_{\sigma\tau\alpha}H_{\mu\nu}H_{\rho}^{\alpha}.
\end{align}
When these requirements are met, the field is determined by
\begin{equation}
    F_{\mu\nu}=\varepsilon_{\mu\nu\alpha}v^{\alpha}
    \quad \text{and} \quad
    v_{\mu}v_{\nu} = \frac{1}{2}H_{\mu\nu}.
\end{equation}

Now, for a null field the conditions are instead
\begin{align}
S_{\alpha\beta}X^{\alpha}X^{\beta} >0, \\
G=-3\Lambda,\\
S_{\mu[\nu}S_{\rho]\sigma}=0,
\end{align}
for some vector field $X$ and
\begin{align}
     S_{\mu\nu}S_{\rho[\sigma;\tau]}+S_{\mu\rho}S_{\nu[\sigma;\tau]}+S_{\nu\rho[;\sigma}S_{\tau]\mu}\notag\\
     =-\mu \varepsilon_{\sigma\tau\alpha}S_{\mu\nu}S_{\rho}^{\alpha}.
\end{align}
Here, $S_{\mu\nu}$ are the components of the traceless Einstein tensor. Moreover, the field is constructed from
\begin{equation}
    F_{\mu\nu}=\varepsilon_{\mu\nu\alpha}v^{\alpha}
    \quad \text{and} \quad
    v_{\mu}v_{\nu} = \frac{1}{2}S_{\mu\nu}.
\end{equation}

These conditions constitute a generalization of the ones provided in Ref.~\cite{Krongos:2016bqp}, which are recovered upon setting $\mu$, the topological mass of the gauge field, to zero. The proof is straightforward and is along the same lines of that work.

\section{Dimensions higher than three}
\label{highD}

The composite version of the gravitational Cheshire effect which we have found in this paper occurs in every dimension; it is not a three-dimensional phenomena. Indeed, the solution provided by Eq. \eqref{PhiM} is unaltered if one adds flat dimensions to the metric in Eqs. \eqref{AnsatzM}. In contrast, the stealth fields overflying the AdS spacetime possesses gauge fields that are always massive. As a consequence, those Maxwell-Chern-Simons fields do not carry over to higher-dimensional versions. Nonetheless, Proca fields may be employed in all dimensions if so desired. The calculations are straightforward and follow from our approach in the previous section. It would be interesting to study configurations such as these with a St\"uckelberg field, thus complementing investigations such as those of Refs.~\cite{Marrani:2017uli,Marrani:2022hva}. 

As mentioned above, higher-dimensional versions of the Sol plane waves have been studied in different gravitational contexts. A non-inheriting solution to Eqs. \eqref{eom}, in arbitrary spacetime dimension $D$, is found over the conformally flat plane-wave spacetime
\begin{equation}
    \rd s ^2 =-2\rd u \rd v+\sum^{D-2}_{i=1} \rd x^2_i+\frac{2x^2_i}{u^2}\rd u^2,
\end{equation}
with Eq. \eqref{nisol}, as is, and Maxwell field given by
\begin{equation}
     f(u)^2=\frac{\Phi^2}{4}\left(\log{\frac{\Phi}{u^{D-2}}}\right)''-\frac{D-2}{u^2}.
\end{equation}

Homogeneous plane waves with this precise time dependence arise as Penrose limits, Refs.~\cite{Penrose:1965rx,Penrose:1976}, of spacetimes near their singularity. This occurs near cosmological and black hole singularities alike, in dimensions greater than or equal to four, see Ref.\cite{Blau:2004yi}.

On a similar note, recall that higher-dimensional versions of the Nil universe have also been studied in gravitational settings. For example, Minkowski and AdS spacetimes emerge as fluctuations on such Heisenberg spaces, as explained in Ref.~\cite{Bravetti:2015ija}. Here, we show that these odd-dimensional manifolds are sourced by constant nonminimally coupled scalar fields and nonnull Maxwell-Chern-Simons fields. To illustrate this point consider the five-dimensional case, where Eq. \eqref{eom} is modified only in that the gauge field must instead comply with
\begin{equation}
    \rd\star F + \mu F\wedge F=0.
\end{equation}

To closely follow the three-dimensional case, we consider a Lorentzian spacetime with Kundt geometry on the five-dimensional Heisenberg group. Such geometries are natural generalizations of five-dimensional pp-waves, as discussed in Ref.~\cite{Coley:2002ku}. The system is composed by
\begin{subequations}
    \begin{align}
    \rd s ^2 &= -2\rd u \rd v+\rd y^2+ \rd z^2 \notag\\
    &\qquad +(\rd x+v\rd u+y\rd z)^2, \\
    \Phi &= \Phi_0, \\
    F &= q\rd u \wedge \rd v+ p\rd y \wedge \rd z.
    \end{align}
\end{subequations}
It requires
\begin{subequations}
\begin{equation}
    \Lambda = m^2 = 0, \quad
    \mu = -\frac{p+q}{2pq},
\end{equation}
together with
\begin{equation}
    \Phi_0=\sqrt{\frac{2(p^2+q^2)+1}{\zeta}}
    \qquad \text{and} \quad
    p^2=q^2.
\end{equation}
\end{subequations}
This five-dimensional solution shows that the appearance of the Chern-Simons term in the three-dimensional case is not a dimensional accident and one should expect their appearance in the higher-dimensional cases. Moreover, it illustrates that in some dimensions the gauge field allows for a massless solution; a Maxwell field. 

In this section, we have presented higher-dimensional pp-waves and generalizations of them whose underlying structure is that of a nilmanifold or a solvmanifold. Such solutions complement the use of these spaces in gravitational physics such as the black hole spacetimes constructed in Refs.~\cite{Hervik:2003vx,Hervik:2007zz}.

\end{document}